\documentclass[preprint]{aastex}
\usepackage{graphicx}
\usepackage{natbib}
\usepackage{epstopdf}
\received{}
\revised{}
\accepted{}

\slugcomment{In preparation for the Astrophysical Journal}

\shortauthors{Wood et al.}
\shorttitle{3D H{\sc ii} Regions}

\begin{document}

\title{Three dimensional geometries and the analysis of H{\sc ii} regions}

\author{Kenneth Wood\altaffilmark{1, 2}, J.~E. Barnes\altaffilmark{1}, Barbara Ercolano\altaffilmark{3}, L.~M. Haffner\altaffilmark{2}, R.~J. Reynolds\altaffilmark{2}, J.~Dale\altaffilmark{4}}

\altaffiltext{1}{School of Physics \& Astronomy, University of St Andrews, 
North Haugh, St Andrews, Fife, KY16 9AD, Scotland; kw25@st-andrews.ac.uk}

\altaffiltext{2}{Department of Astronomy, University of Wisconsin-Madison, 475 N. Charter St., Madison, 
WI 53706, USA}

\altaffiltext{3}{UniversitŠts-Sternwarte Munchen, Scheinerstr. 1, 81679 Munchen, Germany, Boltzmannstrasse 2, 85748 Garching, Germany}

\altaffiltext{4}{Excellence Cluster `Universe', Boltzmannstr. 2, 85748 Garching, Germany}

\authoremail{kw25@st-andrews.ac.uk}

\begin{abstract}
We compare emission line intensities from photoionization models of smooth and fractal shell geometries for low density H{\sc ii} regions, with particular focus on the low-ionization diagnostic diagram [N{\sc ii}]/H$\alpha$ vs H$\alpha$. Building on previously published models and observations of Barnard's Loop, we show that the observed range of intensities and variations in the line intensity ratios may be reproduced with a three dimensional shell geometry. Our models adopt solar abundances throughout the model nebula, in contrast with previous one dimensional modeling which suggested the variations in line intensity ratios could only be reproduced if the heavy element abundances were increased by a factor of $\sim 1.4$. For spatially resolved H{\sc ii} regions, the multiple sightlines that pierce and sample different ionization and temperature conditions within smooth and fractal shells produce a range of line intensities that are easily overlooked if only the total integrated intensities from the entire nebula model are computed. Our conclusion is that inference of H{\sc ii} region properties, such as elemental abundances, via photoionization models of one dimensional geometries must be treated with caution and further tested through three dimensional modeling.

\end{abstract}


\section{Introduction}

Analysis of emission lines from H{\sc ii} regions provides information on the ionization state and temperature of the gas, elemental abundances within the nebula, and the Lyman continuum spectrum of the ionizing source(s). Observations of H{\sc ii} regions are usually interpreted using one-dimensional models where an external source ionizes a slab or an internal source ionizes a sphere or spherical shell and the global averages of the emission line intensities and their ratios are compared with observations \citep[e.g.,][]{peq2001}. Although employing one-dimensional radiation transfer, such models contain a wealth of very detailed physics describing atomic, molecular, and dust processes. 

Recent code developments have extended photoionization modeling to include three dimensional radiation transfer \citep[e.g.,][]{olr98, ercolano2003, wme2004} and thus the ability to study the projected intensities from highly asymmetric nebulae. For example, models by \citet{wood2005} of the H$\alpha$ ($\lambda=6563$~\AA) and [N{\sc ii}] ($\lambda=6583$~\AA) emission lines from the $\zeta$~Oph H{\sc ii} region employing a fractal density distribution demonstrated the variety of line ratios that could be obtained. Ultimately the data and models in this analysis were compared by taking radial averages of the H$\alpha$ intensity and  [N{\sc ii}]/H$\alpha$ line ratio. \citet{ercolano2007} and \citet{ewb2010} explored the effects of a 3D distribution of ionizing sources on abundance determinations. These authors found that a centrally concentrated distribution of ionizing stars results in a higher effective ionization parameter in the nebula, compared to a more spread out distribution of the same stellar population. This can give errors of up to an order of magnitude in abundance, when strong line methods are used. However they found that abundance determinations based on direct measurements of temperature (by use of nebular to auroral line diagnostics) are not affected. Models of star forming regions presented by \citet{ercolano2011} incorporating 3D hydrodynamics and photoionization further demonstrate the variety of line ratios present in spatially resolved 2D images.

There are many uncertainties in photoionization modeling including atomic data, ionizing spectra, elemental abundances, and, the focus of this paper, the geometry of the gas. The two most popular models to describe H{\sc ii} region data are that of a Stromgren sphere of gas ionized by a central source \citep{stromgren1939} and the filling factor model where ionized blobs are surrounded by vacuum \citep[e.g., see discussion and examples in][]{OFAGN2}. Neither of these approaches represents the reality of an H{\sc ii} region containing ionized and neutral gas with a range of temperatures, densities, and ionization states. Smooth shell models do have a range of temperatures and ionization states that change as a function of distance from the ionizing source, but do not have neutral inclusions and ionized-neutral interfaces that naturally arise due to shielding in 3D geometries. Moving to 3D geometries gives a greater variety of temperature and ionization states within a model H{\sc ii} region and a correspondingly larger spread in projected intensities \citep{ercolano2011}.

Recently, using the CLOUDY photoionization code \citep[as described in][]{ferland1998}, \citet{Odell2011} analyzed new and archival emission line data from Barnard's Loop and concluded that there may be an increase in elemental abundances. Their models assumed that the abundances of all elements were increased in unison and as such did not take into account different processes producing the elements. In this paper we construct spherical shell and 3D fractal shell geometries for Barnard's Loop that demonstrate the observations may be reproduced without appealing to abundance variations. 

\section{3D photoionization simulations}

The photoionization models are constructed using the 3D Monte Carlo code described by \citet{wme2004}. The code discretizes the H{\sc ii} region density on a 3D linear Cartesian grid and computes the electron temperature and ionization structure of H, He, C, N, O, Ne, and S, the elements that dominate the cooling in low density H{\sc ii} regions. We do not consider photons with energies above 54~eV, the ionization potential of He$^+$, which is appropriate for the ionizing sources and consistent with observations which demonstrate there is very little ionized helium in low density H{\sc ii} regions and diffuse ionized gas \citep{rt95}. Due to grid resolution, we do not compute emissivities for cells where the neutral fraction is greater than 0.25. Such cells represent ionized-neutral interfaces and our current 3D Cartesian grid cannot spatially resolve the rapid change in ionization state at the interface.

We follow \citet{Odell2011} and start off by constructing spherically symmetric models comprising a central point source illuminating a shell of uniform density, $n = 3\, {\rm cm}^{-3}$, with inner and outer radii $R_{\rm in} =46$~pc and $R_{\rm out} = 55$~pc. 
We fix the ionizing luminosity at $Q=2.5\times 10^{49}\, {\rm s}^{-1}$ as estimated by \citet{Odell2011} for the stars believed to be ionizing Barnard's Loop. Photoionization models are constructed for source temperatures 29000~K and 32000~K, spanning the estimated temperature range of the ionizing stars. The ionizing spectra for our simulations were taken from the WM-BASIC library of \citet{sternberg2003} assuming solar abundances and $\log g=4$. 
For 3D models we use the same inner and outer radii and convert the uniform density shell to a fractal geometry using the algorithm of \citet{elmegreen1997} as described in several of our recent papers \citep[e.g.,][]{mww2002,wood2005}. This algorithm leaves a fraction, $f_{\rm smooth}$, in a smooth, uniform density component, and redistributes the remainder into hierarchical clumps. We adopt $f_{\rm smooth}=1/3$, so the lowest density in our fractal Barnard's Loop models is $n=1\, {\rm cm}^{-3}$.

Images of Barnard's Loop \citep[e.g., Figure~1 in][]{Odell2011} suggest a shell thickness of around 10pc. This is indeed achieved in our simulations because the smooth models are ionization bounded at around 54pc. For the fractal models, their smooth component is ionized out to the outer edge at 55pc, but because of the dominance of ionized higher density clumps the overall appearance is that of a thin fragmented shell (see Figure~1 below).

In all the simulations presented below the elemental abundances are He/H$=0.1$, C/H$=1.4\times 10^{-4}$, 
N/H$=6.5\times 10^{-5}$, O/H$=4.3\times 10^{-4}$, Ne/H$=1.17\times 10^{-4}$, and S/H$=1.4\times 10^{-5}$. Our photoionization models provide emissivities and hence projected intensities for various emission lines and for this paper we compute maps of H$\alpha$ ($\lambda=6563$~\AA) and [N{\sc ii}] ($\lambda=6583$~\AA). Our code also provides the intensity of the two closely spaced lines of [S{\sc ii}] ($\lambda=6716${~\AA} and $\lambda=6731$~\AA). However, as discussed by \citet{Odell2011}, we can shift the model [S{\sc ii}]/H$\alpha$ intensity ratio by an unknown amount because we do not know the dielectronic recombination rates for sulfur. Therefore, in this paper we focus our attention on the [N{\sc ii}]/H$\alpha$ vs H$\alpha$ diagram.

\section{Results}

Figure~1 shows the H$\alpha$ intensity and [N{\sc ii}]/H$\alpha$ line ratio maps for the 29000~K source for the smooth and fractal shell models (the corresponding maps for the 32000~K source are qualitatively similar). We can immediately see the H$\alpha$ limb brightening for the smooth shell and the more fragmented, but still ring-like structure for the fractal shell model. Figures~2 and 3 present the intensity and line ratio maps as scatter plots for the diagnostic diagram [N{\sc ii}]/H$\alpha$ vs H$\alpha$. Clearly a wide range of values occur for both H$\alpha$ intensity and line ratios as we now explain.

The ionizing luminosity, shell density and radial extent, are such that the smooth shell is ionization bounded (i.e., all ionizing photons are absorbed before the outer edge of the shell), resulting in very high temperatures at the edge of the ionized volume. This is shown in Figure~4 and arises because only the highest energy photons can penetrate to the outermost edges of the ionized volume, hence producing the highest temperatures in the nebula. In contrast the fractal shell has a smooth density component which is $n=1\,{\rm cm}^{-3}$ and this component is ionized, unless shadowed by dense clumps. Therefore the temperatures in the smooth component of the fractal model are smaller than for the corresponding radius in the smooth shell model (see Figure~4). Hence there is not the very large [N {\sc ii}]/H$\alpha$ associated with low H$\alpha$ intensities as seen at the edges of the smooth shell. 

The two ionizing source temperatures for the smooth model yield line ratios that nicely bracket the observations. The double valued nature of the [N{\sc ii}]/H$\alpha$ vs H$\alpha$ diagram for the smooth model shown in Figure 2 is explained with reference to Figure~4 which shows intensity cuts and also the radial temperature profile within the shell. Towards the edge of the shell the H$\alpha$ intensity rises and falls with radius due to limb brightening. However, the [N{\sc ii}]/H$\alpha$ ratio increases with radius because sightlines towards the edge of the shell are sampling higher and higher temperatures. Hence the double-valued nature of this diagnostic diagram. The points with low H$\alpha$ and large [N{\sc ii}]/H$\alpha$ arise at large radius while the low H$\alpha$ and [N{\sc ii}]/H$\alpha$ are towards the inner regions of the shell. It is unlikely that such double-valued diagnostic diagrams will be observed because of the difficulty of accurately subtracting foreground and background emission from the warm ionized medium, the low density ionized component of the interstellar medium \citep[see discussion of background effects in][]{wood2005}.

The right hand panels in Figures 2 and 3 show all impact parameters between $46\,{\rm pc}<R<54\,{\rm pc}$, a range appropriate for comparison to Barnards loop.  It is clear from figure 2 that a smooth density structure is unable to match the observations, a result also found by \citet{Odell2011} which led them to suggest abundance variations as a method for better modelling the region.  However figure 3 shows that if a fractal 3D model is used, it is possible to match the observations without the need to alter elemental abundances.  

The fractal shell models presented in Figures~1 and 3 display a wider range of H$\alpha$ intensities due to the density contrasts within the simulation, with densities in the fractal shell often exceeding $n> 30\, {\rm cm}^{-3}$. In addition, the values of [N{\sc ii}]/H$\alpha$  extend to lower values than the uniform density models. This corresponds to the cells in the simulation with the lowest densities ($n=1\, {\rm cm}^{-3}$) being ionized and the corresponding lower temperatures in the smooth ionized component as described above. Figure 4 shows the range of temperatures in the fractal models, nicely explaining the lower [N{\sc ii}]/H$\alpha$ compared to the smooth models. The [N{\sc ii}]/H$\alpha$ vs H$\alpha$ diagram for the 29000~K source (red dots in Figure~3) shows that for all but the lowest H$\alpha$ intensities the line ratio lies in the range 
$0.2\la$ [N{\sc ii}]/H$\alpha \la 0.25$, bracketing the data presented by \citet{Odell2011}. 

Our 3D models do include the high temperatures that occur at ionized-neutral interfaces (see Figure~4). However our Cartesian grid does not adequately spatially resolve these boundaries in the fractal models. In our simulations the percentage of ionized grid cells we ignore with neutral fractions above 0.25 is less than 10\% of the ionized volume. We anticipate that, while important, better resolution of interfaces within the simulation will not change our overall conclusions regarding the effects of 3D geometries on the line ratio diagnostic diagrams of H{\sc ii} regions studied in this paper. The resolution effects of our Cartesian grid do mean that we cannot make accurate predictions of, for example, the [O{\sc i}] ($\lambda=6300$~{\AA}) emission line which is temperature sensitive and through charge exchange is tied to the ionization state of hydrogen. The [O{\sc i}]/H$\alpha$ intensity ratio is larger in diffuse ionized gas than in H{\sc ii} regions, suggesting that high temperatures possibly associated with interface emission is important in widespread diffuse ionized gas \citep[e.g.,][]{hausen2002}.

The results we have presented for analytically produced fractal densities are also seen in photoionization simulations of hydrodynamical simulations. The star forming region models presented by \citet{ercolano2011} clearly demonstrate the same effects of a wide range of observed line ratios in spatially resolved images. See their figures 4 and 5 for a very nice example of diagnostic diagrams where the synthetic line ratios span a range which includes values that are usually associated with shock ionization. However, the \citet{ercolano2011} results are from purely photoionization simulations, again demonstrating that 1D analysis can lead to misunderstandings and misclassifications of the physical processes producing observed emission line ratios.

\section{Summary}

Through 3D photoionization modeling we have demonstrated that the observed line ratios from Barnard's Loop may be reproduced without appealing to increased elemental abundances. We have focussed on the [N{\sc ii}]/H$\alpha$ vs H$\alpha$ diagnostic diagram since accurate predictions of [S{\sc ii}] emission is not possible due to unknown dielectronic recombination rates for sulfur. The simple explanation of our result is that 3D geometries provide a range of temperatures and ionization states and hence a range of emission line intensities different from the total integrated intensity from 1D shell models  \citep[see also][]{ercolano2011}. However, the 1D shell models presented in Figure~2 do in fact demonstrate that it is possible to reproduce the observations with smooth models when rays piercing different impact parameters are considered. A spatially resolved analysis of the projected intensities from smooth models was not presented by \citet{Odell2011} --- this is the main point at which our analyses differ and the reason for our different conclusions regarding elemental abundances within Barnard's Loop. The uncertainties introduced by different atomic databases and 3D geometries make it difficult to determine absolute H{\sc ii} region abundances from analysis of diagnostic diagrams. However, it will be possible to determine relative abundances and abundance gradients from photoionization modeling of H{\sc ii} regions at different spatial locations within a galaxy.

\acknowledgements
We thank Bob O'Dell and Gary Ferland for constructive criticism of our manuscript

\bibliographystyle{apj}
\bibliography{references}

\begin{figure}
\includegraphics{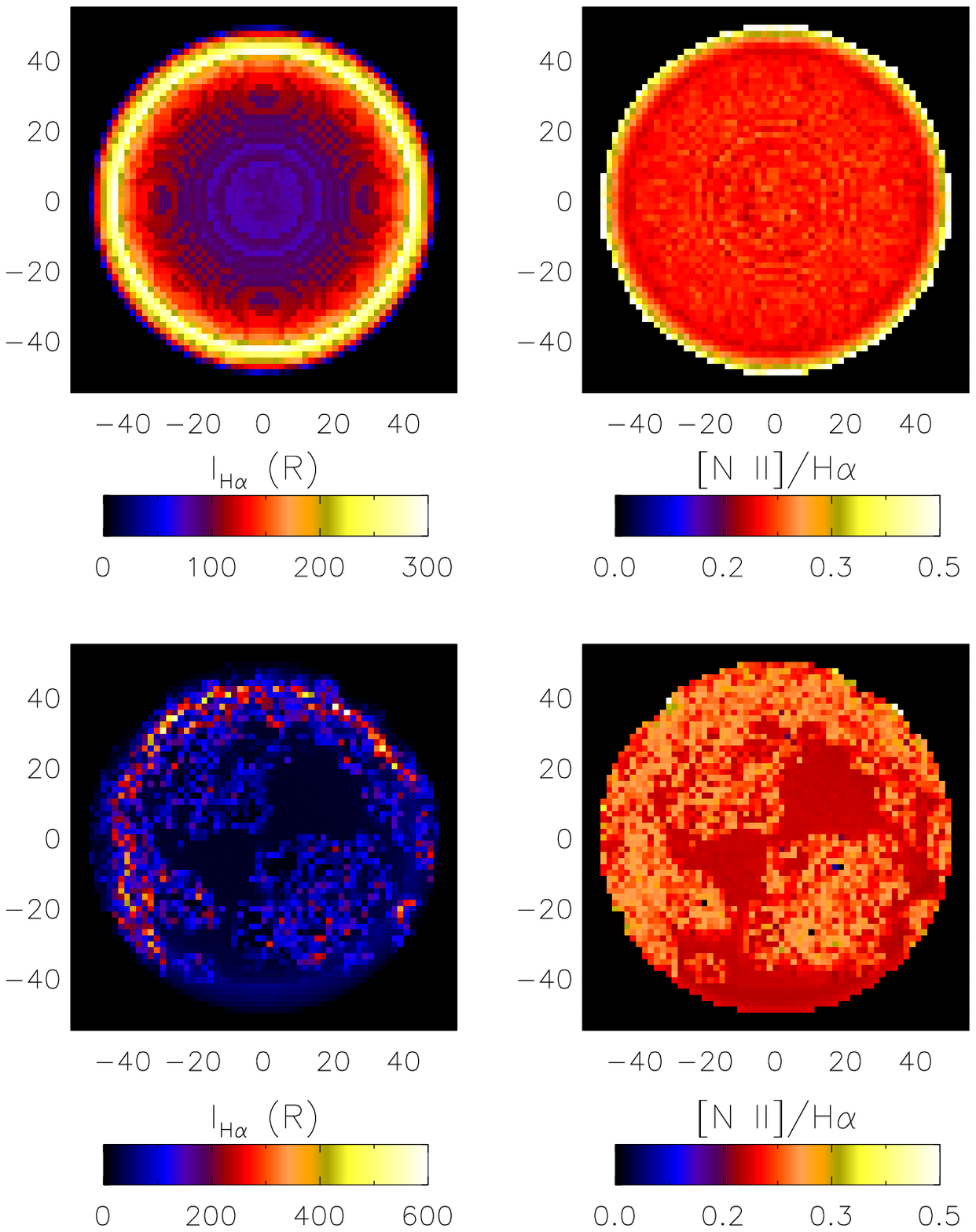}
\caption{H$\alpha$ intensity maps (left) and [N{\sc ii}]/H$\alpha$ line ratio maps (right) for the smooth shell (upper panels) and fractal shell (lower panels) models. The axes are labeled in units of parsecs. The values of the intensities and line ratios are shown as scatter plots in Figures~2 and 3}
\end{figure}

\begin{figure}
\includegraphics[scale=0.6,angle=90]{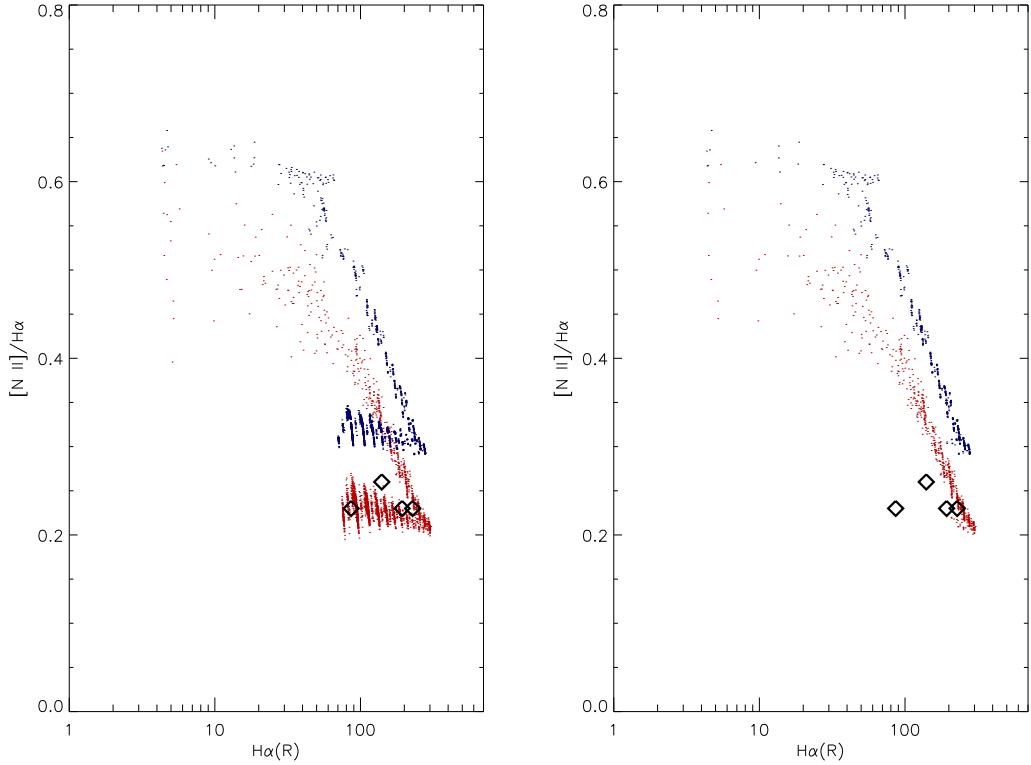}
\caption{Diagnostic diagrams for the smooth shell models. Red and blue dots are results for the 29000~K and 32000~K sources respectively. In the left panel, each dot represents a different sightline through the model nebula, corresponding to the individual pixels in the two dimensional maps presented in Figure~1.  The right panel shows the models for impact parameters in the range  $46\,{\rm pc}<R<54\,{\rm pc}$, corresponding to the observations of Barnard's Loop (diamonds). The double-valued pattern in the left hand panel arises from the increasing temperature with radius in the shell and the corresponding rise then fall of the H$\alpha$ intensity with radius. The highest values for the line ratios occur at the outer edge of the nebula (see Figure~4).}
\end{figure}

\begin{figure}
\includegraphics[scale=0.6,angle=90]{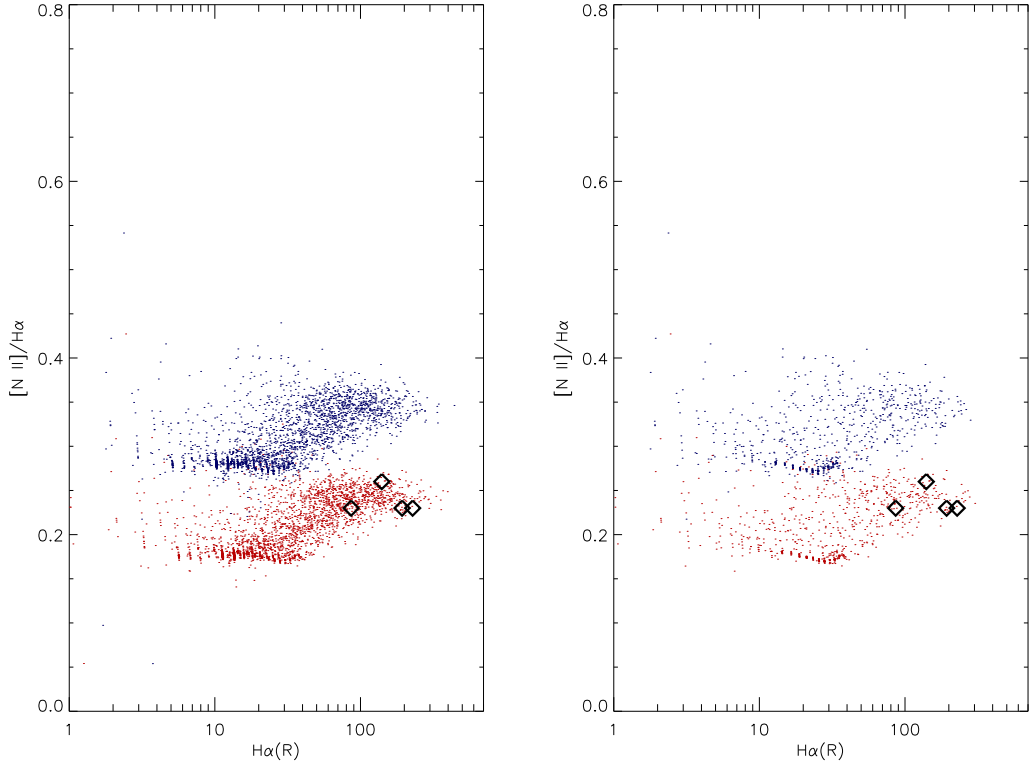}
\caption{As for Figure~2, but for the fractal shell models. Left panel shows intensities and line ratios for all impact parameters $R<55\,{\rm pc}$, right panel restricts impact parameters to be in the range $46\,{\rm pc} < R < 54\,{\rm pc}$. Notice the wide range of H$\alpha$ intensities that arise from the range of densities, temperatures, and ionization fractions sampled by different sightlines through the models. The [N{\sc ii}]/H$\alpha$ values do not reach the high values as for the smooth models because the smooth component of the fractal models is almost fully ionized and therefore at a lower temperature than the corresponding radius in the smooth models.}
\end{figure}

\begin{figure}

\includegraphics[scale=0.6,angle=90]{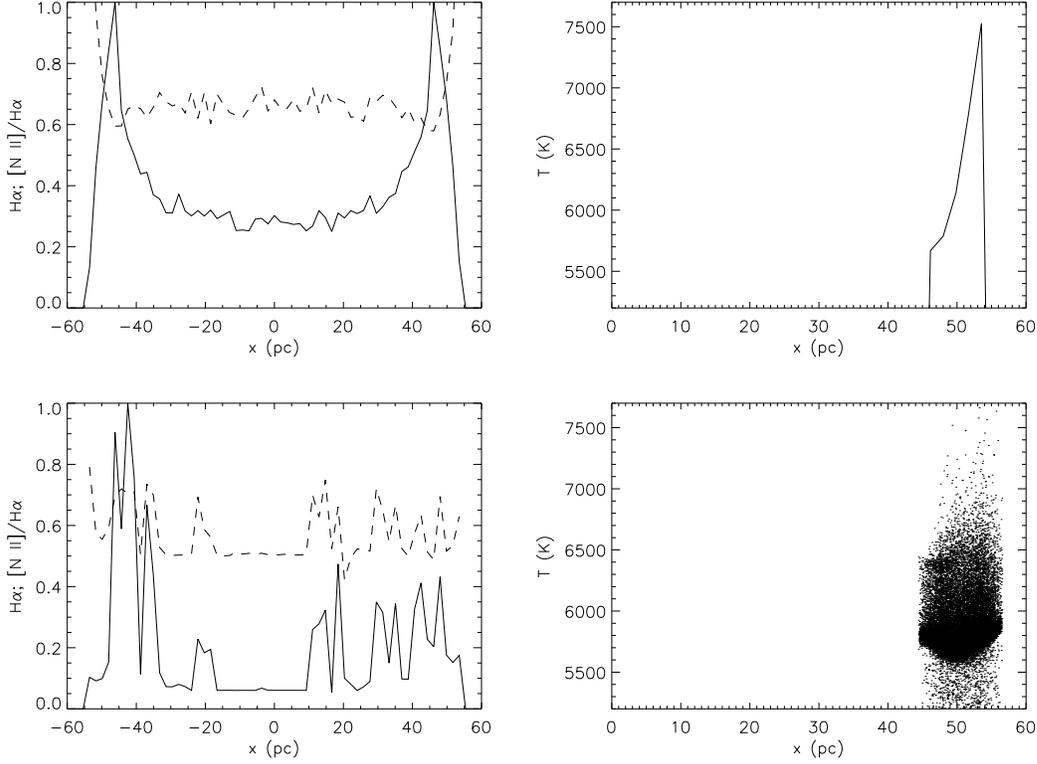}
\caption{Left panels: intensity cut across the center of the H$\alpha$ (solid line) and [N{\sc ii}]/H$\alpha$ (dashed line) line ratio maps from Figure~1 for the smooth (upper) and fractal (lower) models. The intensities and line ratios have been normalized to their maximum values. This explains the double-values obtained in the H$\alpha$ vs [N{\sc ii}]/H$\alpha$ diagnostic diagram for the smooth models in Figure~2. Upper right panel: radial temperature profile of the smooth model showing the rise in temperature towards the edge of the H{\sc ii} region. Lower right panel shows the temperature at all radial points in the fractal shell model. The concentration of points between 5500~K and 6000~K corresponds to the smooth component of the fractal shell --- notice these temperatures are lower at large radii than for the smooth shell model. Although the shell physically extends to $R_{\rm out} = 55$pc, the ionizing luminosity is such that for the smooth shell the edge of the H{\sc ii} region occurs around $r\sim 54$pc, hence the rapid rise in temperature as the gas turns neutral and only highest energy photons can reach the outermost locations. The fractal shell's smooth component is ionized out to the outer edge of the shell (low temperatures) and the high temperatures present in the lower right panel arise at the ionized-neutral interfaces that occur throughout the fractal shell.}

\end{figure}

\end{document}